\documentclass[prl,showpacs,groupedaddress,floatfix,twocolumn]{revtex4}%

\usepackage{amsmath}
\usepackage{amsfonts}
\usepackage{amssymb}
\usepackage{graphicx}%
\usepackage{color}

\begin{document}
\title{On the nature of spatiotemporal light bullets in bulk Kerr media}
\author{D. Majus$^1$, G. Tamo\v{s}auskas$^1$, I. Gra\v{z}ulevi\v{c}i\={u}t\.{e}$^1$, N. Garejev$^1$,
A. Lotti$^2$, A. Couairon$^3$, D. Faccio$^4$, A. Dubietis$^1$}

\affiliation{$^1$Department of Quantum Electronics, Vilnius
University, Saul\.{e}tekio Ave. 9, Building 3, LT-10222 Vilnius,
Lithuania}

\affiliation{$^2$Dipartimento di Scienza e Alta Tecnologia,
Universit\`{a} degli Studi dell'Insubria, Via Valleggio 11,
I-22100 Como, Italy}

\affiliation{$^3$Centre de Physique Th\'{e}orique, CNRS, Ecole
Polytechnique, F-91128 Palaiseau, France}

\affiliation{$^4$School of Engineering \& Physical Sciences,
Heriot-Watt University, Edinburgh, UK}

\date\today

\begin{abstract}
We present a detailed experimental investigation, which uncovers
the nature of light bullets generated from self-focusing in a bulk
dielectric medium with Kerr nonlinearity in the anomalous group
velocity dispersion regime. By high dynamic range measurements of
three-dimensional intensity profiles, we demonstrate that the
light bullets consist of a sharply localized high-intensity core,
which carries the self-compressed pulse and contains approximately
25\% of the total energy, and a ring-shaped spatiotemporal
periphery. \emph{Sub-diffractive} propagation along with
\emph{dispersive broadening} of the light bullets in free space
after they exit the nonlinear medium indicate a strong space-time
coupling within the bullet. This finding is confirmed by
measurements of spatiotemporal energy density flux that exhibits
the same features as stationary, polychromatic Bessel beam, thus
highlighting the physical nature of the light bullets.
\end{abstract}


\maketitle

Propagation-invariant electromagnetic wave-packets -- light
bullets, are long sought in many areas of modern optics and
attract a great deal of interest in fundamental and applied
research \cite{Wise2002}. Generation of three-dimensional light
bullets, which propagate in the medium without natural dispersive
broadening and diffractive spreading, is a non-trivial task from
the analytical and numerical points of view, and even more
complicated to achieve in real experimental settings
\cite{Malomed2005,Mihalache2012,Leblond2013}.

The pursuit for three-dimensional light bullets considers two
essentially different physical concepts. The first is based on the
generation of spatiotemporal solitons, which could be regarded as
ideal light bullets with rapidly (exponentially) decaying tails,
constituting a high degree of energy localization. Formation of
the spatiotemporal soliton relies on simultaneous balancing of
diffraction and dispersion by nonlinear effects, such as
self-focusing and self-phase modulation \cite{Silberberg1990}. In
the first approximation, these conditions could be met in media
with Kerr nonlinearity in the anomalous group velocity dispersion
(GVD) range \cite{Silberberg1990}. However, the proposed light
bullet is a three dimensional extension of the universal Townes
profile, therefore possessing similar properties: it forms only at
the nonlinear focus \cite{Moll2003}, and is modulationally
unstable \cite{Porras2007a}. Therefore, for achieving a
spatiotemporal invariant propagation, linear and nonlinear optical
properties of the medium must be suitably tailored, see e.g.
\cite{Belic2008,Burgess2009,Chen2009,Torner2009,Lobanov2010}, thus
raising difficult technological challenges. So far, to the best of
our knowledge, experimental demonstration of truly three
dimensional light bullets was reported only in a two-dimensional
array of coupled waveguides featuring quasi-instantaneous cubic
nonlinearity and a periodic, transversally modulated refractive
index \cite{Minardi2010}.

The second concept for achieving light bullets is based on the
precise tailoring of the input wave packet so as to match the
material properties, without the need for optical nonlinearities,
i.e. defeating the natural diffractive spreading and dispersive
broadening in the linear propagation regime. Linear light bullets
are non-solitary, weakly localized wave packets, whose stationary
propagation is achieved due to Bessel-like profile of the beam,
whose spectral components are distributed over certain propagation
cones, so as to continuously refill the axial part, which contains
an ultrashort pulse. Moreover, this approach is equally effective
in media with normal, as well as anomalous GVD. To this end,
non-solitary, weakly localized spatiotemporal linear light bullets
have been experimentally demonstrated in the form of the X-waves
\cite{Saari1997}, Airy bullets \cite{Chong2010,Abdollahpour2010}
and ultrashort-pulsed Bessel-like beams \cite{Bock2012}. However,
practical realizations of the linear light bullets require precise
control of propagation angles and phases of the spectral
components, and therefore intricate experimental techniques.

A much more straightforward route for achieving non-solitary,
weakly localized light bullets is based on self-reshaping of the
entire wave-packet (ultrashort pulsed laser beam) by means of
nonlinear effects, producing the nonlinear analogs of the X-waves
\cite{Conti2003,DiTrapani2003} and Airy bullets
\cite{Panagiotopoulos2013}. In particular, spontaneous formation
of the nonlinear X-waves was demonstrated by means of femtosecond
filamentation in transparent dielectric media, where the input
Gaussian-shaped wave packet self-adjusts its spatiotemporal shape
via nonlinear effects into a specific spatiotemporal X shape,
which maintains its stationarity even in the presence of the
nonlinear losses \cite{Kolesik2004,Couairon2006,Faccio2006}.
However, pulse splitting which occurs during filamentation of
intense femtosecond pulses in the normal GVD regime
\cite{Couairon2007}, prevents formation of a single X wave.

Conversely, studies of filamentation in the conditions of
anomalous GVD predicted the generation of isolated spatially and
temporally compressed pulses
\cite{Moll2004,Berge2005,Liu2006,Skupin2006,Chekalin2013,Smetanina2013}.
Recent investigations have uncovered a filamentation regime, in
which quasi-stable three-dimensional non-spreading-pulses are
generated \cite{Durand2013a}. However, there remain relevant
unanswered questions regarding the interpretation of these light
bullets. In this Letter we explicitly demonstrate that the light
bullets generated by self-focusing of femtosecond-pulsed beams in
bulk dielectric media with instantaneous Kerr nonlinearity and
anomalous group velocity dispersion are polychromatic Bessel-like
wave-packets, which bear the basic properties of the nonlinear
O-waves \cite{Porras2005}, as verified by simultaneous
measurements of spatiotemporal profiles, frequency-resolved
angular spectra, near field intensity distributions, linear and
nonlinear propagation features and energy density flux.

The experiment was performed using 90 fs Gaussian pulses with
center wavelength of 1.8 $\mu$m from a commercial, Ti:sapphire
laser system-pumped optical parametric amplifier (Topas C, Light
Conversion Ltd.). Its output beam (an idler wave, in present case)
was suitably attenuated, spatially filtered and focused by an
$f=+100$ mm lens into 45-$\mu$m (FWHM) spot size which was located
on the front face of the nonlinear medium (sapphire sample). The
input pulse energy of 3.1 $\mu$J (corresponding to 3.4 $P_{\rm
cr}$, where $P_{\rm cr}=10$ MW is the critical power for
self-focusing in sapphire) was set so as to induce a light
filament, which formed after 4 mm of propagation, as verified by
the supercontinuum emission in the visible spectral range. The
spatiotemporal intensity distribution at the output of sapphire
sample was measured by three-dimensional imaging technique, based
on recording spatially-resolved cross-correlation function, see
e.g. \cite{Majus2010}. More specifically, the output beam was
imaged (with $5\times$ magnification) onto a 20-$\mu$m-thick
beta-barium borate crystal and gated by means of broadband
sum-frequency generation with a short, 25-fs pulse with central
wavelength of 720 nm from a noncollinear optical parametric
amplifier (Topas-White, Light Conversion Ltd.), which was pumped
by the second harmonic of Ti:sapphire laser system. The
cross-correlation signal with center wavelength of 515 nm was then
imaged (with $1.8\times$ magnification) onto the CCD camera
(Grasshopper 2, Point Grey) with a pixel size of 4.4 $\mu$m and
14-bit dynamic range. By changing the time delay of the gating
pulse in a 8-fs step, we acquired a sequence of cross-correlation
images, which afterwards were merged together to reproduce the
entire spatiotemporal intensity distribution of the light bullet.

\begin{figure}[ht!]
\includegraphics[width=8.5cm]{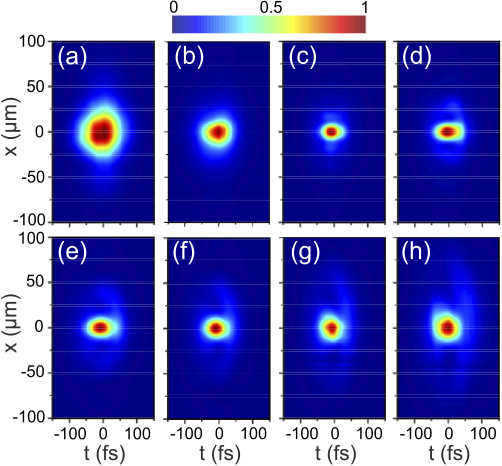} \caption{(Color online)
Spatiotemporal intensity distributions of self-focusing Gaussian
wave-packet, which transforms into a light bullet, as measured at
various propagation lengths $z$ in sapphire: (a) $z=0$ mm, (b)
$z=3.1$ mm, (c) $z=4.2$ mm, (d) $z=6.0$ mm, (e) $z=8.2$ mm, (f)
$z=9.0$ mm, (g) $z=12.3$ mm, (h) $z=15.2$ mm.}\label{fig:xt}
\end{figure}

\begin{figure}[ht!]
\includegraphics[width=8.0cm]{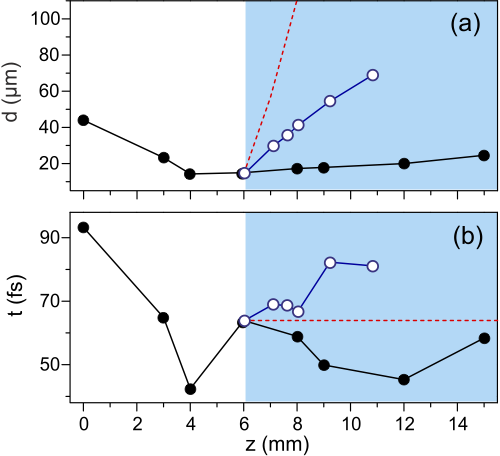} \caption{(Color online)
Evolution of (a) beam FWHM diameter and (b) pulsewidth versus
propagation distance. Full circles show formation and propagation
dynamics of the light bullet in sapphire, as summarized from
Fig.~\ref{fig:xt}. Open circles show propagation of the light
bullet in free space (air), which starts at $z=6$ mm. Dashed
curves indicate the expected free-space propagation features of a
strongly localized Gaussian wave packet.}\label{fig:freespace}
\end{figure}

Evolution of the spatiotemporal intensity distribution over
propagation distance $z$ was captured by placing sapphire samples
of different lengths in such a way, that the output face of the
sample was always kept at the same fixed position, while moving
the focusing lens accordingly, to ensure the location of the input
focal plane at the front face of the sample. Figure~\ref{fig:xt}
presents the spatiotemporal intensity profiles of the wave-packet
as measured at various propagation distances in sapphire, showing
how the input Gaussian wave-packet transforms into a spatially and
temporally compressed three-dimensional light bullet. Over the
first few mm of propagation, the input Gaussian beam shrinks due
to self-focusing and the input Gaussian pulse experiences
self-compression due to the interplay between self-phase
modulation and anomalous GVD. After the nonlinear focus ($z=4.2$
mm), the wave-packet transforms into a spatially and temporally
compressed three-dimensional light bullet, which has FWHM diameter
of 15 $\mu$m and pulse duration of 40 fs, as evaluated from the
deconvolution of the on-axis cross-correlation function. High
dynamic range measurements reveal that the light bullet consists
of a sharply localized high-intensity core, which carries the
self-compressed pulse and a low-intensity, ring shaped
spatiotemporal periphery, and propagates without an apparent
change of its spatiotemporal shape. Propagation dynamics of the
light bullet are summarized in Fig.~\ref{fig:freespace}, where
full circles show the beam width and pulse duration versus
propagation distance, demonstrating that high intensity core
maintains its localization over more than 10 mm of propagation,
that exceeds 25 Rayleigh ranges, as calculated for the Gaussian
beam of equivalent dimensions.

In a further experiment, where the light bullet after 6 mm of
propagation in sapphire was thereafter let to propagate in free
space (air), we recorded interesting and very important
propagation features. We observe gradual increase of both, spatial
and, more importantly, temporal dimensions of the central core, as
the propagation distance in free space increases, as shown by open
circles in Fig.~\ref{fig:freespace}. These data are compared with
the calculated linear evolution of a strongly localized Gaussian
wave-packet with identical spatial and temporal dimensions (shown
by the dashed curves), which represent the expected spreading of a
soliton-like object in free space. The distinctive differences in
free space propagation between the present light bullet and a
soliton-like object are immediately clear: the diffraction
spreading of the bullet is almost 5 times less than that of a
Gaussian-shaped beam, and its temporal width increases by a factor
of 1.3, just after 3 mm of propagation, in the absence of
dispersion, while for a soliton-like object it is expected to
remain constant. These results demonstrate that the light bullets,
as they exit the nonlinear medium, exhibit \emph{sub-diffractive}
and \emph{dispersive} propagation in free space, that is
incompatible with the behavior of highly localized spatiotemporal
soliton-like objects. The linear and nonlinear propagation
features of the light bullets arise from dramatic spatiotemporal
reshaping of the input Gaussian wave-packet and resulting strong
space-time coupling, which is a distinctive property of
conically-shaped wave packets \cite{Faccio2008}.

\begin{figure}[ht!]
\includegraphics[width=8.4cm]{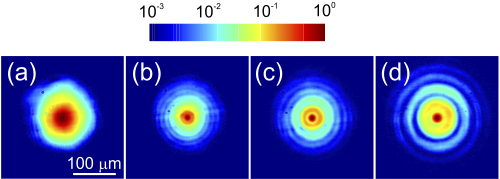} \caption{(Color online)
Spatial profiles of (a) the input beam at $z=0$ mm and the light
bullet at (c) $z=4.2$ mm, (d) $z=6.0$ mm, (e) $z=9.0$
mm.}\label{fig:space}
\end{figure}

In the support of this claim, we studied the relevant
characteristics of the light bullet in more detail.
Figure~\ref{fig:space} illustrates the spatial profiles of the
input Gaussian beam and the light bullet, as recorded at different
propagation distances in sapphire, and obtained by time
integration of spatiotemporal profiles presented in
Fig.~\ref{fig:xt}. The spatial profiles are presented in
logarithmic intensity scale and reveal an intense central core,
which contains approximately 25\% of the total energy and is
surrounded by a low-intensity concentric ring-shaped periphery,
thus resembling a distinct Bessel-like intensity distribution,
which emerges from the interplay of self-focusing, nonlinear
absorption and diffraction \cite{Dubietis2004a}. A Bessel-like
intensity distribution of the light bullet explains the spatial
features of the central core observed in the experiment: its
robustness during the nonlinear propagation in sapphire, as well
as sub-diffractive propagation in free space, which are achieved
via energy refilling from the beam periphery
\cite{Dubietis2004a,Dubietis2004b}.

\begin{figure}[ht!]
\includegraphics[width=8.4cm]{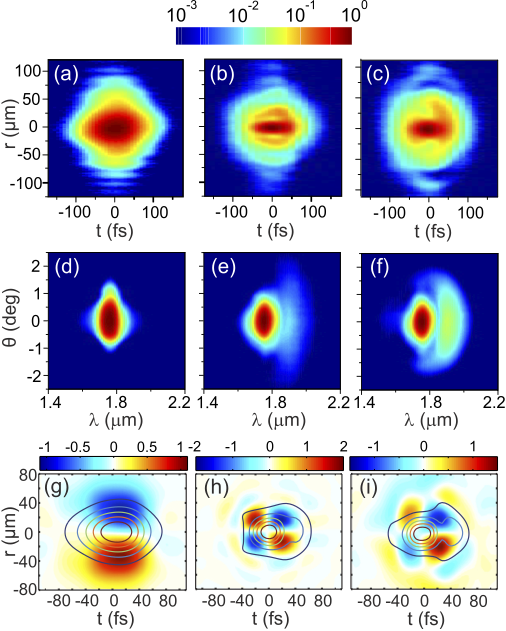} \caption{(Color online)
Spatiotemporal intensity profiles (shown in logarithmic intensity
scale) of (a) the input wave packet, and the light bullet at (b)
$z=6.0$ mm, (c) $z=9.0$ mm. (d), (e) and (f) show the
corresponding angularly resolved spectra. (g), (h) and (i) show
the corresponding retrieved transverse energy fluxes, see text for
details.}\label{fig:bullet}
\end{figure}

Figure~\ref{fig:bullet} highlights the characteristic
spatiotemporal properties of the input wave packet and the light
bullet. Figs.~\ref{fig:bullet}(a)-(c) compare the near-field
($r,t$) spatiotemporal intensity profiles in logarithmic intensity
scale, so as to better visualize the entire spatiotemporal
structure of the light bullet. Figures~\ref{fig:bullet}(d)-(f)
present the corresponding angularly resolved ($\theta,\lambda$)
spectra, as measured by scanning the far-field (at 25 cm distance
from the exit face of sapphire sample) with a 200-$\mu$m fiber tip
of a fiber spectrometer (AvaSpec-NIR256-2.5, Avantes). The
angularly resolved spectra suggest the occurrence of an
elliptical, or O-shaped, pattern of conical emission, that is
expected from Kerr-driven spatiotemporal instability gain profile
in the case of anomalous GVD \cite{Porras2005b,Porras2007}.

The near and far field intensity profiles are combined together to
obtain the full spatio-temporal phase profile of the light bullets
by means of an iterative Gerchberg-Saxton retrieval algorithm, see
\cite{Faccio2009} for details. The gradient of the retrieved phase
profile is used to explicitly visualize the transverse energy
density flux in the full spatial and temporal coordinates, as
shown normalized in Figs.~\ref{fig:bullet}(g), (h) and (i), along
with overlayed contour plot of the retrieved intensity profiles.
Here the color coding is the same as in \cite{Lotti2010}, where
blue and red colors indicate downward and upward fluxes with
respect to the vertical axis, respectively.

\begin{figure}[ht!]
\includegraphics[width=8.0cm]{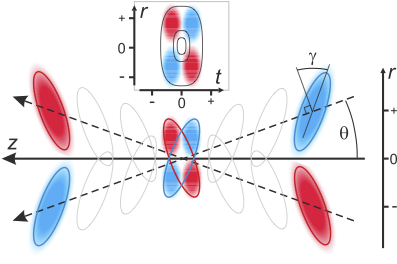} \caption{(Color online)
Schematic representation of transverse energy flux in
ultrashort-pulsed (polychromatic) Bessel beam with subluminally
propagating envelope peak. $\theta$ and $\gamma$ are arbitrary
cone and pulse-font tilt angles,
respectively.}\label{fig:geometry}
\end{figure}

The transverse energy density flux of the light bullet
[Fig.~\ref{fig:bullet}(h), (i)] indicates a radially symmetric
pulse front tilt resulting from strong space-time coupling, and
the spatiotemporal distribution of the currents is almost
identical to that of a stationary, polychromatic Bessel beam with
subluminally propagating envelope peak \cite{Lotti2010}, as
schematically depicted in Fig.~\ref{fig:geometry}. The established
subluminal propagation of the envelope peak is very much in line
with the results of numerical simulations in fused silica
\cite{Durand2013a,Durand2013b}, where the position of the peak is
shown to continuously shift toward positive times with
propagation. The radially symmetric pulse-front tilt, which owes
to angular dispersion \cite{Porras2003}, explains the observed
features of temporal behavior of the light bullet in the nonlinear
and linear (free space) propagation regimes, as illustrated
Fig.~\ref{fig:freespace}(b). The propagation angles of the
spectral components comprising the light bullet compensate for
material dispersion in dispersive medium, whereas such angular
distribution causes the dispersive spreading of the pulse, as the
bullet exits the dispersive medium and propagates in free space.
In addition, presence of the pulse-front tilt may readily explain
the reported pulsewidth dependence on the aperture size in the
autocorrelation measurements \cite{Smetanina2013}. We underline
that a soliton or soliton-like wave-packet would be expected to
exhibit a flat phase and hence no transverse energy density flow
in the spatiotemporal domain. Conversely, the conical flux
unveiled here is inherent to polychromatic Bessel-like pulses and
hence serves as unambiguous demonstration of the nature of the
light bullets generated by self-focusing in a bulk dielectric
media with anomalous GVD. More precisely, the entirety of
established properties of the bullet: quasistationary O-shaped
spatiotemporal intensity profile and characteristic
angularly-resolved spectrum, Bessel-like spatial intensity
distribution and transverse energy flux along with nonlinear and
linear (free-space) propagation features closely resemble those of
the nonlinear O-waves featuring weak losses and subluminally
propagating envelope peak \cite{Porras2005}.

In conclusion, we uncovered the nature of three-dimensional light
bullets generated from self-focusing of intense femtosecond pulses
in bulk dielectric media with anomalous GVD. Self-focusing
dynamics of 100 fs pulses at center wavelength of 1.8 $\mu$m in
sapphire was experimentally captured in detail in full
four-dimensional (x,y,z,t) space by means of three-dimensional
imaging technique. We demonstrate that the emerging
three-dimensional light bullets consist of a sharply localized
high-intensity core, which carries the self-compressed pulse and a
weak, delocalized low-intensity periphery, comprising a
Bessel-like beam. We explicitly demonstrate that the seemingly
weak periphery as viewed in linear intensity scale, is nonetheless
an important integral part of the overall wave packet, as it
continuously balances energy losses in the central core and
prevents it from spreading during its linear and nonlinear
propagation. We disclose that spatiotemporal reshaping of the
input Gaussian wave packet results in development of a very
distinct spatiotemporal flow of the energy that is not compatible
with a spatiotemporal soliton, but rather finds a natural
explanation in terms of a polychromatic Bessel beam, which could
be qualified as a nonlinear O-wave which has weak losses and
subluminally propagating envelope peak \cite{Porras2005}. As a
consequence of this, the light bullets exhibit a rather remarkable
behavior as they exit the sample and propagate in free space
(air), i.e. in the absence of any nonlinear or dispersive effects:
the bullets disperse temporally, yet continue to propagate with
strongly suppressed diffraction. We expect that these important
features are characteristic to an entire family of spatiotemporal
light bullets, which are generated by femtosecond filamentation in
bulk dielectric media with anomalous GVD and should be carefully
accounted for building the basis for diverse future applications
that may require a simple setup for creating intense, temporally
compressed and sub-diffractive wave packets \cite{Hemmer2013}.

This research was funded by the European Social Fund under the
Global Grant measure, grant No. VP1-3.1-\v{S}MM-07-K-03-001. D.F.
acknowledges financial support from the European Research Council
under the European Union's Seventh Framework Programme
(FP/2007-2013)/ERC GA 306559.

\end{document}